\documentclass[12pt,a4paper]{JHEP3}
\usepackage{amsmath,amssymb,bm}
\usepackage{latexsym}
\usepackage{graphicx}

\def\eqref#1{Eq.~(\ref{#1})}

\def\Eq#1{\begin{equation} #1 \end{equation}}
\def\Eqr#1{\begin{eqnarray} #1 \end{eqnarray}}
\def\Eqrsubl#1#2{\begin{subequations}\label{#1}\Eqr{#2}\end{subequations}}

\newcommand{\nn}{\nonumber}
\newcommand{\pd}{\partial}

\def\Xsp{{\rm X}}
\def\bMsp{\bar{{\rm M}}}
\def\Ysp{{\rm Y}}
\def\Zsp{{\rm Z}}
\def\Msp{{\rm M}}
\def\X5sp{{\rm X}_5}
\def\Y3sp{{\rm Y}_3}
\def\Z3sp{{\rm Z}_3}

\def\lap{{\triangle}}
\def\e{{\rm e}}

\title{Warped de Sitter compactifications}
\author{
Masato Minamitsuji\\
Yukawa Institute for Theoretical Physics,
Kyoto University\\
~~Kyoto 606-8502, Japan.\\
~~~E-mail: \email{masato$``$at$"$yukawa.kyoto-u.ac.jp},\\
Department of Physics, Graduate School of Science and Technology\\
~~Kwansei Gakuin University, Sanda 669-1337, Japan.\\
~~~E-mail: \email{masato.minamitsuji$``$at$"$kwansei.ac.jp}}
\author{
Kunihito Uzawa\\
Department of Physics, Kinki University\\
~~Higashi-Osaka, Osaka 577-8502, Japan.\\
~~~E-mail: \email{uzawa$``$at$"$phys.kindai.ac.jp}}

\abstract{%
We show that the warped de Sitter compactifications 
are possible under certain conditions in $D$-dimensional gravitational 
theory coupled to a dilaton, a form field strength, and 
a cosmological constant. 
We find that the solutions of field equations give  
de Sitter spacetime with the warped structure, 
and 
discuss 
cosmological models
directly obtained from these solutions.
We also construct a 
cosmological model in the lower-dimensional effective 
theory.
If there is a field strength having non-vanishing components
along the internal space, the 
moduli can be fixed 
at the minimum of the effective potential
where a de Sitter vacuum can be obtained. 
}
\preprint{YITP-11-108}

\begin{document}


\section{Introduction}
 \label{sec:introduction}

de Sitter compactification of higher-dimensional theory is 
an important cosmological issue.
Such a solution has been explored from several points of view 
because this provides a fairly direct 
explanation of the accelerating expansion of four-dimensional universe. 
The inflation and accelerating expansion \cite{Riess:1998cb, 
Perlmutter:1998np, Riess:2001gk, Riess:2006fw, Kowalski:2008ez, tsujikawa} 
with warped compactifications can 
be constructed in a variety of ways from higher-dimensional cosmology or 
string theory \cite{Kachru:2003aw, Kachru:2003sx, Silverstein:2007ac}. 
Actually, 
the proposal for a physical construction of de Sitter 
compactification has been made recently \cite{
Silverstein:2007ac, Danielsson:2009ff, Caviezel:2009tu, Wrase:2010ew, 
Danielsson:2010bc, Blaback:2010sj, Rosseel:2006fs, Minamitsuji:2010cm}. 
In the gravity theory, 
an initial clue of the de Sitter compactification was that 
the hyperbolic space associated to 
a higher-dimensional theory
can be regarded as internal space 
\cite{Emparan:2003gg, Townsend:2003fx, Ohta:2003pu, Ohta:2003ie, 
Maldacena:2000mw, Chen:2003ij, Chen:2003dca, Chen:2006ia} 
(See also \cite{Neupane:2005nb, Neupane:2006in, Neupane:2009jn, 
Neupane:2010is, Neupane:2010ey}). 
It is not hard to see why a de Sitter compactification 
give an the hyperbolic space that can be derived from higher-dimensional 
Einstein equations. We can see, roughly speaking, that a curvature of 
the de Sitter space can be compensated by that of the internal space.
Then, the curvature of the internal space has the negative sign. 

The warped de Sitter compactification shed much light on whether 
there was the exact solution of field equations because 
the solution such as D-branes or M-branes in the supergravity 
can be embedded in warped compactification.
Thus the cosmological solutions of the $p$-brane system have been 
discussed in ten- or eleven-dimensional supergravity theory as the 
examples of warped compactifications 
\cite{Binetruy:2007tu, Maeda:2009zi, Minamitsuji:2010kb, 
Gibbons:2005rt, Chen:2005jp, Kodama:2005fz, Kodama:2005cz, 
 Maeda:2009tq, Gibbons:2009dr, Maeda:2009ds, Maeda:2010yk, Maeda:2010ja}. 
These solutions are very efficient for constructing the cosmological 
models, and showing that they indeed are compactifications with 
warped structure \cite{Kodama:2005fz, Kodama:2005cz, Lu:1995cs}.
However, such constructions cannot make the accelerating 
expansion of our four-dimensional universe manifest yet 
\cite{Minamitsuji:2010fp, Maeda:2010aj, Minamitsuji:2010uz, 
Maki:1992tq}.

A road to obtain the warped de Sitter compactification has 
appeared recently in a study of
the higher-dimensional pure gravity theory in \cite{Neupane:2010ya}. 
It was shown that 
warped structure of the spacetime realizes
a de Sitter universe,
but 
the pay is that one of the internal space dimensions 
has to have an infinite volume rather than small and finite one. 
For such a class of solutions, 
a way of the construction of a cosmological model
is to insert a brane world boundary in the noncompact direction
whose world volume contains all the remaining compact directions 
of the internal space 
which is discussed more explicitly in \cite{Neupane:2010ya}
(see also 
\cite{Minamitsuji:2010kb, Minamitsuji:2010fp, Minamitsuji:2010uz}).
It may be regarded as a generalization of
the five-dimensional brane world models (see e.g., \cite{rs}).
The details of the brane world models obtained
from our solutions are argued in \cite{Minamitsuji:2011gn}.

In the first part of the paper, we will focus on
the derivation of the new warped de Sitter solutions
obtained as the generalization of Ref. \cite{Neupane:2010ya}.
We begin with the pure gravity 
theory in $D$ dimensions, and re-examine
possible generalizations of the de Sitter 
compactification \cite{Neupane:2010ya}, 
with the internal space of positive and negative curvature. 
This gives the logically simple treatment of this topic
and 
the clearest explanation of warped compactifications. 
In the case with a positively curved internal space,
we will show a simpler expression of the solution \cite{Neupane:2010ya}. 
We then discuss the warped de Sitter compactification with
several matter fields,
in particular 
a scalar and gauge fields in higher-dimensional gravity theory 
because of the following two reasons.

One is that in general
the solutions with matter fields are not 
given 
simply
via the trivial extension of the pure gravity solutions
\cite{Neupane:2010ya}. 
How the matter fields
backreact on the pure gravity solution
is a highly nontrivial question.
Even though the spacetime structures
with and without matter have
similarities in some regions,
in other regions
they lead to different pictures.
An instructive example
is the case of static and spherically symmetric
black holes in four dimensions,
namely Schwarzschild black hole (without the gauge field)
and Reissner-Nordstr$\ddot{{\rm o}}$m 
black hole (with the gauge field).
Although both solutions have
asymptotically flat infinities if there is no cosmological constant,
the inner structure of these black holes
can be quite different.
A similar but more relevant example to us
is the case of the time-dependent D3-brane solutions in the ten-dimensional 
type IIB supergravity with the trivial dilaton
\cite{Binetruy:2007tu, Gibbons:2005rt}. 
In the presence of the 5-form 
field strength, 
the asymptotic spacetime structure behaves as a Kasner spacetime
and contains the horizons of black $3$-branes,
while in the absence of it
the whole spacetime structure is exactly 
described by a Kasner spacetime with no horizons.
Therefore,
whether (and how) the matter field,
in particular, the gauge field 
gives some differences in the spacetime structure
for the warped de Sitter solutions
should be carefully studied. 
Moreover, 
a variety of phenomenologically interesting objects
in the higher-dimensional theories
arise in the presence of matter fields.
For example,  
D-branes in string theory arise 
if the gravity is coupled to several combinations
of scalars and forms.

The other 
reason is to achieve
the stabilization of the internal space modulus
by changing properties of the internal space,
the matter fields or other details. 
Such an analysis 
can be particularly meaningful
by adding gauge fields \cite{Carroll:2001ih},
and cannot be achieved in the 
previous papers \cite{Neupane:2010ya}
without the gauge field.
The importance of the gauge field
can be seen at the level of the cosmological model,
for example to obtain the well-behaved effective four-dimensional Planck mass
in the limit of the vanishing expansion rate
\cite{Minamitsuji:2011gn}.
Along the way, we
will clarify some physical properties of warped de Sitter 
compactifications.

In the second part of this paper, 
we will investigate a stabilization
mechanism of the internal space via some kinds of matter fields
in the $D$-dimensional theory,
and present another way to construct a cosmological model
adding matter fields
after integrating over the compact directions and 
involving the noncompact direction in the higher-dimensional 
solution into our Universe.
We consider the matter fields with a cosmological constant 
in order to stabilize the scale of internal
space. Many works suggest that the field strength might provide a physical
mechanism which is capable of accounting for the extreme
smallness of the extra dimensions 
\cite{Maeda:1985bq, Maeda:1985gz, Uzawa:2003ji, Uzawa:2003qh}.

The present paper is constructed as follows.
We give the warped de Sitter
compactifications in $D$-dimensional pure gravity theory
in Sec.~\ref{sec:D-dim}.
As the starting point, 
we will investigate the $D$-dimensional solutions 
in the pure gravity
which are given in terms of the warped product
of the $(n+1)$-dimensional external spacetime $\Msp$ and
the $(D-n-1)$-dimensional internal space $\Zsp$.  
A de Sitter compactification is
obtained if we specialize to the case 
that the $(n+1)$-dimensional spacetime $\Msp$
is given by the product of 
$\Msp=\Xsp\times\mathbb{R}$ with the warped structure, 
where X is being an $n$-dimensional de Sitter spacetime.
We then present the solutions to be warped compactification 
with several matter fields, and this will also lead to $n$-dimensional 
de Sitter space in Sec.~\ref{sec:m}. 
We then briefly discuss the brane world models
where the $n$-dimensional $\Xsp$ space
becomes our de Sitter Universe,
following \cite{Minamitsuji:2011gn}.
In Sec.~\ref{sec:ef},
we also discuss the stabilization of moduli degrees of freedom in the 
lower-dimensional effective theories 
for the warped compactification. 
We present a 
construction of the cosmological model
in which
a cosmological $(n+1)$-dimensional spacetime $\bMsp$
is obtained from a compactification of the $D$-dimensional theory
onto a $(D-n-1)$-dimensional Einstein space.
Here $\bMsp$ is conformally related to the
$(n+1)$-dimensional metric $\Msp$. 
After the dimensional reduction,
the moduli degrees of freedom are coupled 
to the matter fields in the external spacetime. 
Upon reducing on the internal space, 
the $(n+1)$-dimensional effective 
action would lead to the moduli potential 
which could have a local minimum of the potential,
where the Einstein frame manifold $\bMsp$
becomes our $(n+1)$-dimensional de Sitter universe. 
Sec. \ref{sec:Discussions} is devoted to 
giving concluding remarks.

\section{De Sitter spacetime in $D$-dimensional warped compactifications}
\label{sec:D-dim}

In this section, we construct the cosmological  
warped compactification which gives the accelerating expansion of the universe 
in the pure gravity.

We take the following ansatz for the $D$-dimensional metric
\Eq{
ds^2=\e^{2A_1(y)}\left[\e^{2A_0(x)}q_{\mu\nu}(\Xsp)dx^{\mu}dx^{\nu}
+u_{ij}(\Ysp)dy^idy^j\right],
 \label{v:metric:Eq}
}
where $q_{\mu\nu}$ is a $n$-dimensional metric which
depends only on the $n$-dimensional coordinates $x^{\mu}$, 
and $u_{ij}$ is the $(D-n)$-dimensional metric which
depends only on the $(D-n)$-dimensional coordinates $y^i$. 

In terms of the $D$-dimensional metric (\ref{v:metric:Eq}),  
the Einstein equations are written as
\Eqrsubl{v:Einstein:Eq}{
&&
R_{\mu\nu}(\Xsp)-(n-2)D_{\mu}D_{\nu}A_0
+(n-2)\pd_{\mu}A_0\pd_{\nu}A_0
-q_{\mu\nu}\left[\lap_{\Xsp}A_0
+(n-2)q^{\rho\sigma}\pd_{\rho}A_0
\pd_{\sigma}A_0\right]\nn\\
&&\hspace{1cm}-\e^{2A_0}q_{\mu\nu}\left[
\lap_{\Ysp}A_1+(D-2)u^{kl}\pd_kA_1\pd_lA_1\right]=0,
 \label{v:Einstein-mn:Eq}\\
&&R_{ij}(\Ysp)-(D-2)D_iD_jA_1+(D-2)\pd_iA_1\pd_jA_1\nn\\
&&\hspace{1cm}
-u_{ij}\left[\lap_{\Ysp}A_1+(D-2)u^{kl}\pd_kA_1\pd_lA_1\right]
=0\,,
 \label{v:Einstein-ij:Eq}
}
where $D_{\mu}$, $D_i$ are the covariant derivatives with respective to 
the metric $q_{\mu\nu}$, $u_{ij}$, and  
$\triangle_{\Xsp}$ and $\triangle_{\Ysp}$ are 
the Laplace operators on the space of 
${\rm \Xsp}$ and the space ${\rm \Ysp}$, and 
$R_{\mu\nu}(\Xsp)$ and $R_{ij}(\Ysp)$ are the Ricci tensors
of the metrics $q_{\mu\nu}$ and $u_{ij}$, respectively.

In the following, we consider the warped de Sitter compactifications 
with three kinds of the internal manifold.

\subsection{Internal space with positive curvature}
Let us consider the vacuum solution with the internal space which 
has the positive curvature \cite{Neupane:2010ya}.  
We see the solution whose $(D-n)$-dimensional metric has the form
\Eq{
u_{ij}(\Ysp)dy^idy^j=\rho^2\left[G(y)dy^2+E\gamma_{ab}(\Zsp)dz^adz^b\right]\,,
   \label{p:metric:Eq}
}
where $G(y)$ and $E$ are given by \cite{Neupane:2010ya}
\Eq{
E=\frac{1}{3}(D-n-2),~~~~
G(y)=\frac{1}{12}(D-n+2)\left(\pd_y
A_1\right)^2\,.
  \label{p:metric2:Eq}
}
Though this $(D-n)$-dimensional metric 
has been already present 
in \cite{Neupane:2010ya}, we discuss the dynamical solution beyond 
what has already been cited. 
In terms of the metric of the internal space \eqref{p:metric2:Eq}, 
the Einstein equations \eqref{v:Einstein:Eq} are given by
\Eqr{
&&R_{\mu\nu}(\Xsp)-(n-2)D_{\mu}D_{\nu}A_0
+(n-2)\pd_{\mu}A_0\pd_{\nu}A_0\nn\\
&&\hspace{1cm}
-q_{\mu\nu}\left[\lap_{\Xsp}A_0
+(n-2)q^{\rho\sigma}\pd_{\rho}A_0
\pd_{\sigma}A_0\right]-c_v\,\e^{2A_0}q_{\mu\nu}=0\,,
 \label{p:Einstein-mn:Eq}\\
&&R_{ab}(\Zsp)-4(D-n-2)(D-n+2)^{-1}(D-2)\gamma_{ab}(\Zsp)=0\,,
 \label{p:Einstein-ij:Eq}
}
where $R_{ab}(\Zsp)$ is the Ricci tensor with respect to the 
metric $\gamma_{ab}(\Zsp)$, and $c_v$ is defined by 
\Eq{
c_v\equiv\frac{12(D-2)}{\rho^2(D-n+2)}\,.
   \label{p:c:Eq}
}
Suppose that the function $A_0$ is such that $A_0=A_0(t)$. 
We set the $n$-dimensional spacetime metric $q_{\mu\nu}$ 
\Eq{
q_{\mu\nu}(\Xsp)dx^{\mu}dx^{\nu}
  =-dt^2+a^2(t)\delta_{mn}dx^{m}dx^{n}\,,
    \label{p:n-metric:Eq}
}
where $\delta_{mn}$ is the metric of 
$(n-1)$-dimensional Euclidean space. 
Form the Eq.~(\ref{p:Einstein-mn:Eq}), we find
\Eqrsubl{p:cEinstein2:Eq}{
&&\hspace{-1.2cm}(n-1)\left[\left(\frac{\dot{a}}{a}\right)^{\dot{}}
+\left(\frac{\dot{a}}{a}\right)^2+\ddot{A}_0+\dot{A}_0^2\right]
-c_v\,\e^{2A_0}=0,
    \label{p:cEinstein2-tt:Eq}\\
&&\hspace{-1.2cm}\left[\left(\frac{\dot{a}}{a}\right)^{\dot{}}
+(n-1)\left(\frac{\dot{a}}{a}\right)^2+\ddot{A}_0
+(2n-3)\frac{\dot{a}}{a}\,\dot{A}_0
+(n-2)\dot{A}_0^2-c_v\,\e^{2A_0}\right]q_{mn}=0,
   \label{p:cEinstein2-ab:Eq}
}
where $\dot{}$ denotes the ordinary derivative 
with respect to the coordinate $t$\,. 
We assume that 
the functions $A_0(t)$ and $a(t)$ are given by 
\Eq{
\dot{A}_0=\xi_1\,\e^{A_0}\,,~~~~~~\frac{\dot{a}}{a}=\xi_2\,\e^{A_0}\,,
  \label{p:ansatz:Eq}
}
where $\xi_1$ and $\xi_2$ are constants. 
Thus, the Einstein equations \eqref{p:cEinstein2:Eq} 
are reduced to 
\Eq{
\left[(n-1)\left(\xi_1+\xi_2\right)^2-c_v\right]\,
\e^{2A_0}\,q_{\mu\nu}=0\,.
    \label{p:cEinstein-mn3:Eq}
}
In order to satisfy \eqref{p:cEinstein-mn3:Eq}, the constant 
$c$ obeys 
\Eq{
c_v=(n-1)\left(\xi_1+\xi_2\right)^2\,.
  \label{p:c2:Eq}
}
Form the Eq.~(\ref{p:ansatz:Eq}), we find 
\Eq{
A_0(t)=-\ln\left[-\left(\xi_1t+\xi_3\right)\right]\,,~~~~
a(t)=a_0\left(\xi_1t+\xi_3\right)^{-\xi_2/\xi_1}\,,
   \label{p:ansatz for Aa:Eq}
}
where $\xi_3$ and $a_0$ are constants.
Then, the metric of $D$-dimensional spacetime can be written as
\Eq{
ds^2=\e^{2A_1(y)}\left[-d\tau^2
+a_0^2\,\e^{2H\tau}\delta_{mn}dx^{m}dx^{n}
+u_{ij}(\Ysp)dy^idy^j\right]\,,
  \label{p:D-metric2:Eq}
}
where the cosmic time $\tau$ and the Hubble parameter $H$ are given by 
\Eqrsubl{p:parameter:Eq}{
\tau&=&\pm\frac{1}{\xi_1}\ln\left[-\left(\xi_1t+\xi_3\right)\right]\,,
    \label{p:ctime:Eq}\\
H&\equiv&\frac{d\ln a}{d\tau}=-\left(\xi_1+\xi_2\right)
=\pm\sqrt{\frac{c_v}{n-1}}\,.  
    \label{p:Hubble:Eq}  
}
If we consider the case   
\Eqr{
\xi_1+\xi_2=-\sqrt{\frac{c_v}{n-1}}\,,
   \label{p:Lambda:Eq}
}
the solution leads to an accelerating 
expansion of the $n$-dimensional spacetime. 

In terms of the ansatz \eqref{p:ansatz:Eq}, 
the condition $A_0=0$ corresponds to $\xi_1=0$\,. 
Then, the solution for $a(t)$ is given by 
\Eq{
a(t)=\exp\left(\pm\sqrt{\frac{c_v}{n-1}}\, t\right)\,,
   \label{p:solution for a:Eq}
}
where we used the definition \eqref{p:c:Eq}. 

Clearly, in this case, we can choose $A_0(t)=0$ without loss of generality 
because $A_0(t)$ is the degree of freedom corresponding to the scale of time.
Up to an inessential scaling, the solution of the Einstein equations 
\eqref{v:Einstein:Eq} can be written by
\Eqr{
ds^2=\e^{2A_1(y)}\left[-dt^2+\e^{2Ht}
\delta_{mn}dx^{m}dx^{n}
+u_{ij}(\Ysp)dy^idy^j\right],
  \label{p:D-metric:Eq}
}
where we used the Hubble parameter
\Eq{
H^2=\frac{c_v}{n-1}\,.
  \label{p:Hubble2:Eq}
}
As the Hubble parameter $H$ is proportional to the constant $c_v$\,, 
the internal manifold becomes flat space in the limit 
$H\rightarrow 0\,$.

The warp factor $A_1(y)$ is determined by integrating 
\eqref{p:metric2:Eq} after
specifying the function $G(y)$.
Choosing the function $G(y)=\rho^{-2}$,
it can be described in a much simpler way
than in \cite{Neupane:2010ya}.
Then, integrating \eqref{p:metric2:Eq}
the warped factor becomes
\Eq{
A_1(y)
=-\frac{2}{\rho}\sqrt{\frac{3}{D-n+2}}(y-y_0)
=-\sqrt{\frac{n-1}{D-2}}H(y-y_0),
}
where we have used 
\eqref{p:c:Eq}
and
\eqref{p:Hubble2:Eq}.

\subsection{Internal space with negative curvature}
Next we consider a solution which has the internal space with 
negative curvature. 
We assume that the $(D-n)$-dimensional metric 
takes the form
\Eq{
u_{ij}(\Ysp)dy^idy^j=dy^2+\e^{2\sigma y}\delta_{ab}dz^adz^b\,,
  \label{n:metric:Eq}
}
where $\sigma$ is constant
and $\delta_{ab}$ denotes the metric of 
the $(D-n-1)$-dimensional torus. 
The Einstein equations \eqref{v:Einstein-ij:Eq} then reduce to
\Eqrsubl{n:Einstein:Eq}{
&&\hspace{-1cm}(D-2)\pd_y^2A_1+(D-n-1)\sigma\left(\pd_yA_1+\sigma\right)=0\,,
  \label{n:Einstein y:Eq}\\
&&\hspace{-1cm}\left[\pd_y^2A_1+\left\{(D-n-1)\sigma+(D-2)\pd_yA_1\right\}
\left(\pd_yA_1+\sigma\right)\right]u_{ab}=0\,.
 \label{n:Einstein ab:Eq}
 }
These equations are satisfied if the function $A_1$ is given by 
\Eq{
A_1=-\sigma (y-y_0)\,,
   \label{n:solution A1:Eq}
}
where $y_0$ is constant. 
In terms of \eqref{n:solution A1:Eq}, 
\eqref{v:Einstein-mn:Eq} leads to 
\Eqr{
&&R_{\mu\nu}(\Xsp)-(n-2)D_{\mu}D_{\nu}A_0
+(n-2)\pd_{\mu}A_0\pd_{\nu}A_0\nn\\
&&\hspace{1cm}
-q_{\mu\nu}\left[\lap_{\Xsp}A_0
+(n-2)q^{\rho\sigma}\pd_{\rho}A_0
\pd_{\sigma}A_0\right]-\chi\,\e^{2A_0}q_{\mu\nu}=0\,,
 \label{n:cEinstein-mn3:Eq}
}
where the constant $\chi$ is defined by 
\Eq{
\chi\equiv\lap_{\Ysp}A_1+(D-2)u^{kl}\pd_kA_1\pd_lA_1=(n-1)\sigma^2\,.
   \label{n:chi:Eq}
}
We set the function $A_0=0$ from the beginning
and choose the $n$-dimensional spacetime metric 
$q_{\mu\nu}$ as follows: 
\Eqr{
&&q_{\mu\nu}(\Xsp)dx^{\mu}dx^{\nu}
  =-dt^2+a^2(t)\delta_{mn}dx^{m}dx^{n}\,,
    \label{n:n-metric:Eq}
}
where $\delta_{mn}$ is the metric of 
$(n-1)$-dimensional Euclidean space. 

Upon setting 
(\ref{n:n-metric:Eq}), 
the Einstein equations 
(\ref{v:Einstein-mn:Eq}) are given by 
\Eqrsubl{n:cEinstein-mn4:Eq}{
&&\hspace{-1.9cm}\left[(n-1)\left\{\left(\frac{\dot{a}}{a}\right)^{\dot{}}
+\left(\frac{\dot{a}}{a}\right)^2
\right\}-\chi
\right]q_{tt}=0,
    \label{n:cEinstein-mn41:Eq}\\
&&\hspace{-1.9cm}
\left[\left(\frac{\dot{a}}{a}\right)^{\dot{}}
+(n-1)\left(\frac{\dot{a}}{a}\right)^2
-\chi
\right]q_{mn}=0\,.
   \label{n:cEinstein-mn42:Eq}
}
They give $\left(\frac{\dot{a}}{a}\right)^{\dot{}}=0$,
and 
the Hubble parameter $H$ are defined as
\Eqr{
H&\equiv&\frac{d\ln a}{dt}
=\pm\sqrt{\frac{\chi}{n-1}}
=\pm \sigma
\,.  
    \label{n:Hubble:Eq}  
}
We 
find
the $D$-dimensional metric to be 
\Eqr{
ds^2=\e^{-2
\sigma (y-y_0)}\left(-dt^2+\e^{\pm 2\sigma t}
\delta_{mn}dx^{m}dx^{n}+dy^2\right)
+\e^{2\sigma y_0}\,\delta_{ab}\,dz^adz^b\,,
  \label{n:d-metric:Eq}
}
where the metric $\delta_{ab}$ denotes the $(D-n-1)$-dimensional torus. 
The metric \eqref{n:d-metric:Eq}
takes the same form as the $D$-dimensional Minkowski
spacetime.
However, 
our solution is the product of 
the $n$-dimensional de Sitter spacetime and the internal space of 
$\mathbb{R}\times$T${}^{(D-n-1)}$, 
which is
different from the trivial Minkowski spacetime. 
Since the internal space except for the $\mathbb{R}$ direction
has a finite volume, 
this solution represents a de Sitter warped compactification.

\section{Compactifications with matter fields}
\label{sec:m}

In this section, we discuss the warped de Sitter compactification 
which includes several fields and the cosmological constant. 

\subsection{Compactifications with scalar field}
\label{sec:s}
Let us first consider the warped de Sitter compactification 
which includes the metric $g_{MN}$, the scalar fields $\phi$, $\varphi$
the cosmological constant $\Lambda$,
to see how the ansatz strictly restricts the possibility of 
de Sitter compactifications \cite{Minamitsuji:2010cm}.
The action we consider in the Einstein frame is given by 
\Eq{
S=\frac{1}{2\kappa^2}\int \left[\left(R-2\e^{\alpha_{\phi}\phi}\Lambda\right)
\ast{\bf 1} -\frac{1}{2}d\phi\wedge\ast d\phi
-\frac{1}{2}d\varphi\wedge\ast d\varphi\right],
\label{s:action:Eq}
}
where $\kappa^2$ is the $D$-dimensional 
gravitational constant, $\ast$ is the Hodge
operator in the $D$-dimensional spacetime, and $\alpha_{\phi}$ is defined 
by 
\Eq{
\alpha_{\phi}=\sqrt{\frac{2}{D-2}+c}\,,
   \label{s:beta:Eq} 
}
with the constant $c$.

After varying the action \eqref{s:action:Eq} with respect to the metric
and 
the dilaton, we obtain the field equations
\Eqrsubl{s:field equations:Eq}{
&&R_{MN}=\frac{1}{2}\pd_M\varphi\pd_N\varphi+\frac{1}{2}\pd_M\phi\pd_N\phi
+\frac{2}{D-2}\e^{\alpha_{\phi}\phi}\Lambda g_{MN},
   \label{s:Einstein:Eq}\\
&&d\ast d\phi-2\alpha_{\phi}\e^{\alpha_{\phi}\phi}\Lambda\ast{\bf 1}=0\,,
   \label{s:scalar:Eq}\\
&&d\ast d\varphi=0.
   \label{s:scalar2:Eq}
}
To solve the field equations, we assume that the 
$D$-dimensional metric takes the form
\Eq{
ds^2=\e^{2A_1(y)}\left[\e^{2A_0(x)}q_{\mu\nu}(\Xsp)dx^{\mu}dx^{\nu}
+u_{ij}(\Ysp)dy^idy^j\right],
 \label{s:metric:Eq}
}
where $q_{\mu\nu}$ is a $n$-dimensional metric which depends only on 
the $n$-dimensional coordinates $x^{\mu}$, and $u_{ij}$ is the 
$(D-n)$-dimensional metric which depends only on
the $(D-n)$-dimensional coordinates $y^i$.
Furthermore, we assume that the scalar fields $\phi$ and $\varphi$ 
are given by
\Eq{
\phi=-\frac{2}{\alpha_{\phi}}A_1(y)\,,~~~~~\varphi=\varphi_c A_0(x)\,,
    \label{s:ansatz:Eq}
}
where $\varphi_c$ is a constant.

Let us first consider the Einstein equations \eqref{s:Einstein:Eq}. 
Using the assumptions \eqref{s:metric:Eq} and \eqref{s:ansatz:Eq}, the
Einstein equations are given by
\Eqrsubl{s:Eisntein2:Eq}{
&&\hspace{-1.2cm}
R_{\mu\nu}(\Xsp)-(n-2)D_{\mu}D_{\nu}A_0
+\left(n-2-\frac{\varphi_c^2}{2}\right)\pd_{\mu}A_0\pd_{\nu}A_0
-q_{\mu\nu}\left[\lap_{\Xsp}A_0\right.\nn\\
&&\hspace{-1.2cm}\left.+(n-2)q^{\rho\sigma}\pd_{\rho}A_0\pd_{\sigma}A_0
+\e^{2A_0}\left\{
\lap_{\Ysp}A_1+(D-2)u^{kl}\pd_kA_1\pd_lA_1+\frac{2}{D-2}\Lambda\right\}
\right]=0,
 \label{s:Einstein-mn2:Eq}\\
&&\hspace{-1.2cm}
R_{ij}(\Ysp)-(D-2)D_iD_jA_1+\left(D-2-\frac{2}{\alpha_{\phi}^2}\right)
\pd_iA_1\pd_jA_1\nn\\
&&\hspace{0.cm}
-u_{ij}\left[\lap_{\Ysp}A_1+(D-2)u^{kl}\pd_kA_1\pd_lA_1
+\frac{2}{D-2}\Lambda\right]
=0\,,
 \label{s:Einstein-ij2:Eq}
}
where $D_{\mu}$, $D_{i}$ are the covariant derivatives with respect to 
the metrics $q_{\mu\nu}(\Xsp)$, $u_{ij}(\Ysp)$, and 
$\triangle_{\Xsp}$, $\triangle_{\Ysp}$ are the
Laplace operators on the space of X and the space Y, 
and $R_{\mu\nu}(\Xsp)$, $R_{ij}(\Ysp)$ are the Ricci
tensors of the metrics $q_{\mu\nu}(\Xsp)$, $u_{ij}(\Ysp)$, 
respectively. 

Let us next consider the scalar field equation. Substituting 
\eqref{s:ansatz:Eq} into Eqs.~(\ref{s:scalar:Eq}) and 
(\ref{s:scalar2:Eq}), we obtain
\Eqrsubl{s:scalar eq:Eq}{
&&\frac{2}{\alpha_{\phi}}\e^{-2A_1}\left[\lap_{\Ysp}A_1
+(D-2)u^{kl}\pd_kA_1\pd_lA_1+\alpha_{\phi}^2\Lambda\right]=0,
   \label{s:scalar11:Eq} \\
&&\varphi_c\,\e^{-2A_0}\left[\lap_{\Xsp}A_0
+(n-2)q^{\rho\sigma}\pd_{\rho}A_0\pd_{\sigma}A_0\right]=0\,.
   \label{s:scalar12:Eq} 
 }
Combining \eqref{s:Eisntein2:Eq} with \eqref{s:scalar eq:Eq}, we get 
\Eqrsubl{s:field:Eq}{
&&
R_{\mu\nu}(\Xsp)-(n-2)D_{\mu}D_{\nu}A_0
+\left(n-2-\frac{\varphi_c^2}{2}\right)\pd_{\mu}A_0\pd_{\nu}A_0\nn\\
&&~~~~-\left[\lap_{\Xsp}A_0
+(n-2)q^{\rho\sigma}\pd_{\rho}A_0
\pd_{\sigma}A_0-c\Lambda\e^{2A_0}\right]q_{\mu\nu}=0,
 \label{s:Einstein-mn3:Eq}\\
&&R_{ij}(\Ysp)-(D-2)D_iD_jA_1+\left(D-2-\frac{2}{\alpha_{\phi}^2}\right)
\pd_iA_1\pd_jA_1+c\Lambda u_{ij}=0,
 \label{s:Einstein-ij3:Eq}\\
&&\lap_{\Ysp}A_1
+(D-2)u^{kl}\pd_kA_1\pd_lA_1+\left(\frac{2}{D-2}+c\right)\Lambda=0
   \label{s:scalar21:Eq}\,,\\
&&\varphi_c\left[
\lap_{\Xsp}A_0+(n-2)q^{\rho\sigma}\pd_{\rho}A_0\pd_{\sigma}A_0\right]=0\,.
   \label{s:scalar22:Eq}
   }
Now we assume the metric in which
\Eqrsubl{s:D-qu-metric:Eq}{
q_{\mu\nu}(\Xsp)dx^{\mu}dx^{\nu}
&=&-dt^2+a^2(t)\delta_{mn}dx^{m}dx^{n}\,.
    \label{s:d-metric:Eq} \\
u_{ij}(\Ysp)dy^idy^j&=&dy^2+\gamma_{ab}(\Zsp)dz^adz^b\,,
     \label{s:D-d-metric:Eq}
}
where 
$\gamma_{ab}(\Zsp)$ is the metric of $(D-n-1)$-dimensional 
Einstein space. Let us consider the Eqs.~(\ref{s:Einstein-ij3:Eq}), 
(\ref{s:scalar21:Eq}). 
If we set $A_1=A_1(y)$, Eqs.~(\ref{s:Einstein-ij3:Eq}) and  
(\ref{s:scalar21:Eq}) are reduced to   
\Eq{
A_1=k(y-y_0)\,,~~~~~~
R_{ab}(\Zsp)=-c\Lambda\gamma_{ab}(\Zsp)\,,
}
where $k\equiv\pm\sqrt{-[c(D-2)+2]\Lambda}/(D-2)$, and $y_0$ is constant, 
and $R_{ab}(\Zsp)$ are the Ricci
tensor of the metric $\gamma_{ab}(\Zsp)$\,. The expression $k$ gives 
$\Lambda<0$. 

Next we consider the Eqs.~(\ref{s:Einstein-mn3:Eq}) and 
(\ref{s:scalar22:Eq}). 
Upon setting, $A_0=A_0(t)$, the field equations 
\eqref{s:Einstein-mn3:Eq} give
\Eqrsubl{s:cEinstein-mn4:Eq}{
&&\hspace{-1.5cm}(n-1)\left[\left(\frac{\dot{a}}{a}\right)^{\dot{}}
+\left(\frac{\dot{a}}{a}\right)^2+
\frac{\dot{a}}{a}\,\dot{A}_0+\ddot{A}_0\right]+
\frac{\varphi_c^2}{2}\,\dot{A}_0^2+c\Lambda\e^{2A_0}=0,
    \label{s:cEinstein-mn41:Eq}\\
&&\hspace{-1.5cm}
\left[\left(\frac{\dot{a}}{a}\right)^{\dot{}}
+(n-1)\left(\frac{\dot{a}}{a}\right)^2
+(2n-3)\frac{\dot{a}}{a}\,\dot{A}_0
+\ddot{A}_0+(n-2)\,\dot{A}_0^2
+c\Lambda\e^{2A_0}
\right]q_{mn}=0,
   \label{s:cEinstein-mn42:Eq}\\
&&\hspace{-1.5cm}\varphi_c
\left[\ddot{A}_0+(n-1)\frac{\dot{a}}{a}\,\dot{A}_0
 +(n-2)\,\dot{A}_0^2\right]=0\,,
   \label{s:scalar2-22:Eq}
}
where $\dot{}$ 
denotes the ordinary derivative with respect to the coordinate $t$\,. 
In the following, we look for the solution in the case $c=0$ or 
$\varphi_c=0$ because it is not so easy to find the solution 
analytically if both $c$ and $\varphi_c$ have non-zero values.

Let us first consider the case of $c=0$, which corresponds to 
$R_{ij}(\Ysp)=0$. 
From the \eqref{s:cEinstein-mn4:Eq}, we get 
\Eq{
A_0(t)=\tilde{c}_1 t+\tilde{c}_3,~~~~a(t)=\tilde{c}_4\,\e^{\tilde{c}_2t}\,,
}
where $\tilde{c}_i~(i=1, \cdots , 4)$ are constants, and $\tilde{c}_1$, 
$\varphi_c$ are 
given by
\Eq{
\tilde{c}_1=-\frac{n-1}{n-2}\tilde{c}_2\,,~~~~~
\varphi_c^2=\frac{2(n-2)}{n-1}\,.
}
If we define the cosmic time $\tau$ by 
$\tau=\tilde{c}_1^{-1}\e^{\tilde{c}_1t+\tilde{c}_3}$, 
the scale factor of $n$-dimensional spacetime can be written as 
\Eq{
a(\tau)=a_0\,\tau^{1/(n-1)}\,,
}
where the constant $a_0$ is given by 
$a_0=\tilde{c}_1^{1/(n-1)}\tilde{c}_4\e^{(n-2)\tilde{c}_3/(n-1)}$\,.
The $D$-dimensional metric thus becomes
\Eqr{
ds^2=\e^{2A_1(y)}\left[-d\tau^2+\tau^{2/(n-1)}
\delta_{mn}dx^{m}dx^{n}
+dy^2+\gamma_{ab}(\Zsp)dz^adz^b\right],
  \label{s:d-metric2:Eq}
}
where we absorbed the inessential scaling into the metric. 
The solution implies that the power of the scale factor for $n=4$ is too small 
to give a realistic expansion law such as that in the matter 
dominated era ($a\propto\tau^{2/3}$) or that in the
radiation dominated era ($a\propto\tau^{1/2}$).

Next we consider the case of $\varphi_c=0$. 
Since the function $A_0(x)$ again describes the conformal rescaling of 
the $n$-dimensional metric, the function $A_0$ can take the constant value 
without loss of generality.  
Thus, from the Eq.~(\ref{s:Einstein-mn3:Eq}), we see  
that $n$-dimensional spacetime is an Einstein space.  
If $c$ vanishes, we can no longer find any de Sitter compactifications.  
The contribution of the parameter $c$ is actually important  
and played a major role in our solution. 
The field equations \eqref{s:cEinstein-mn4:Eq} give
\Eq{
a(t)=\tilde{c}_0\,\e^{\pm\sqrt{-c\Lambda/(n-1)}\,t}\,,
}
where $\tilde{c}_0$ is constant. 
Here we define
the cosmic time $\tau$ and Hubble parameter $H$ as
\Eq{
H^2=-\frac{c}{(n-1)}\Lambda\,.
   \label{s:Hubble:Eq}
}
Equation (\ref{s:Hubble:Eq}) denotes that $R_{ij}(\Ysp)$ should be positive 
so that $H^2>0$ unless $c\Lambda=0$.
We choose 
the metric in which
\Eqrsubl{bw:metric:Eq}{
q_{\mu\nu}(\Xsp)dx^{\mu}dx^{\nu}&=&
-dt^2+\tilde{c}_0^2\,\e^{2H t}\delta_{mn}dx^{m}dx^{n}\,,
    \label{bw3:d-metric:Eq}\\
u_{ij}(\Ysp)dy^idy^j&=&\e^{-2A_1}dy^2
+E_s \gamma_{ab}(\Zsp)dz^adz^b,
   \label{sD:metric:Eq}
}
where $E_s$ is constant, and 
$\gamma_{ab}$ is the metric of $(D-n-1)$-dimensional Einstein space. 
Using the metric \eqref{bw:metric:Eq}, the Einstein equations lead to
\Eqr{
A_1(y)&=&\ln \left[\tilde{c}_1(y-y_0)\right]\,,~~~~~~
\quad
E_s=\frac{D-n-2}{(n-1)H^2}\,,
   \label{bew:solution:Eq}
}
where $y_0$ is constant, and $\tilde{c}_1$ is given by 
\Eq{
\tilde{c}_1^2=\left[1+\frac{2}{(D-2)c}\right]\frac{(n-1)H^2}{D-2}\,.
}
Hence 
the $D$-dimensional metric and the scalar field can be expressed as
\Eqrsubl{st:solution:Eq}{
ds^2
&=&\left[\tilde{c}_1 (y-y_0)\right]^2\left[
-dt^2+\tilde{c}_0^2\,\e^{2Ht}
\delta_{mn}dx^{m}dx^{n}\right]
+dy^2
\nonumber\\
& &+\frac{D-n-2}{(n-1)H^2}\left[\tilde{c}_1 (y-y_0)\right]^2
\gamma_{ab}(\Zsp)dz^a dz^b\,,\\
\phi(y)&=&-\frac{2}{\alpha_{\phi}}\ln\left[\tilde{c}_1 (y-y_0)\right]\,.
}
Note that there is a curvature singularity at $y=y_0$.
If we choose $c$ so that $c=c_{\ast}:=\frac{2(D-n-2)}{n(D-2)}$, 
the internal space becomes a flat spacetime.
Otherwise, 
although the internal manifold is 
topologically flat,
it contains a deficit or surplus solid angle.
A deficit solid angle is obtained for $c>c_{\ast}$,
while a surplus one is done for $c<c_{\ast}$.

\subsection{Compactifications with field strength}
\label{sec:f}
We now 
obtain the solution by generalizing the matter field  
beyond the scalar field system that we have considered so far.
As an example,
let us next consider the warped de Sitter compactification 
which include the metric $g_{MN}$, the scalar field $\phi$, and 
the field strength $F$. 

The action we consider in the Einstein frame is given by 
\Eq{
S=\frac{1}{2\kappa^2}\int \left[\left\{R
-2\e^{-\alpha\phi/(p-1)}\Lambda\right\}\ast{\bf 1}
-\frac{1}{2}d\phi\wedge\ast d\phi
-\frac{1}{2\cdot p!}\e^{\alpha\phi}F \wedge\ast F\right],
\label{fs:action:Eq}
}
where $\kappa^2$ is the $D$-dimensional 
gravitational constant, $\ast$ is the Hodge
operator in the $D$-dimensional spacetime, and $\phi$ is 
the scalar field, and $F$ is $p$-form field strength, 
and $\Lambda$, $\alpha$ are constants.

The $D$-dimensional action \eqref{fs:action:Eq} gives the field equations
\Eqrsubl{fs:field equations:Eq}{
&&R_{MN}=\frac{2}{D-2}\e^{-\alpha\phi/(p-1)}\Lambda\,g_{MN}
+\frac{1}{2}\pd_M\phi\pd_N\phi\nn\\
&&~~~~~~~~~~+\frac{1}{2\cdot p!}\e^{\alpha\phi}
\left(pF_{MA_1\cdots A_{p-1}}{F_N}^{A_1\cdots A_{p-1}} 
-\frac{p-1}{D-2}g_{MN}F^2\right),
   \label{fs:Einstein:Eq}\\
&&d\ast d\phi-\frac{\alpha}{2\cdot p!}\e^{\alpha\phi}F\wedge\ast F
 +\frac{2\alpha}{p-1}\e^{-\alpha\phi/(p-1)}\Lambda\ast{\bf 1}=0\,,\\
   \label{fs:scalar:Eq}
&&d\left(\e^{\alpha\phi}\ast F\right)=0.
   \label{fs:strength:Eq}
}
We adopt the following ansatz for the $D$-dimensional metric:
\Eq{
ds^2=\e^{2A(y)}\left[q_{\mu\nu}(\Xsp)dx^{\mu}dx^{\nu}+dy^2
+\gamma_{ab}(\Zsp)dz^adz^b\right],
 \label{fs:metric:Eq}
}
where $q_{\mu\nu}$ is the $n$-dimensional metric which depends only on 
the $n$-dimensional coordinates $x^{\mu}$, and $\gamma_{ab}$ is the 
$p(=D-n-1)$-dimensional metric which depends only on
the $p$-dimensional coordinates $z^a$.
As for the scalar field and the $p$-form field strength, we take
\Eqrsubl{fs:fields:Eq}{
\phi&=&\frac{2}{\alpha}(p-1)A(y)\,,
    \label{fs:scalar2:Eq}\\
F&=&f\,\Omega(\Zsp)\,,
    \label{fs:strength2:Eq}
}
where $f$ is a constant, and $\Omega(\Zsp)$ is the volume form of the Z space
\Eq{
\Omega(\Zsp)=\sqrt{\gamma}dz^1\wedge\cdots\wedge dz^p\,.
}
Here $\gamma$ denotes the determinant of the metric $\gamma_{ab}(\Zsp)$. 

Let us first consider the Einstein \eqref{fs:Einstein:Eq}. 
Using the assumptions (\ref{fs:metric:Eq}) and (\ref{fs:fields:Eq}), 
the Einstein equations \eqref{fs:Einstein:Eq} are written by 
\Eqrsubl{fs:Einstein eq:Eq}{
&&R_{\mu\nu}(\Xsp)-q_{\mu\nu}\left[A''+(D-2)\left(A'\right)^2
+\frac{2}{D-2}\Lambda-\frac{p-1}{2(D-2)}f^2\right]=0\,,
   \label{fs:Einstein-mn:Eq}\\
&&(D-1)A''+2(p-1)^2\alpha^{-2}\left(A'\right)^2
+\frac{2}{D-2}\Lambda-\frac{p-1}{2(D-2)}f^2=0\,,
   \label{fs:Einstein-yy:Eq}\\
&&R_{ab}(\Zsp)-\gamma_{ab}\left[A''+(D-2)\left(A'\right)^2
+\frac{2}{D-2}\Lambda+\frac{n}{2(D-2)}f^2\right]=0\,,
   \label{fs:Einstein-ab:Eq}
}
where $'$ denotes the ordinary derivative with respect to the coordinate $y$, 
 and $R_{\mu\nu}(\Xsp)$ and $R_{ab}(\Zsp)$ are the Ricci 
tensors of the metrics $q_{\mu\nu}$ and $\gamma_{ab}$, respectively.

Let us next consider the gauge field. Under the assumption 
\eqref{fs:strength:Eq}, the Bianchi identity is automatically satisfied.
Also the equation of motion for the gauge field becomes
\Eqr{
d\left[f\,\e^{(D-2)A}\right]\wedge\Omega(\Xsp)\wedge dy=0\,,
 \label{fs:strength3:Eq}
}
where $\Omega(\Xsp)$ is the volume form of the X space
\Eq{
\Omega(\Xsp)=\sqrt{-q}\,dx^0\wedge\cdots\wedge dx^{n-1}\,.
}
Here $q$ denotes the determinant of the metric $q_{\mu\nu}(\Xsp)$. 
Thus the gauge field equation is automatically
satisfied under the assumption \eqref{fs:fields:Eq}\,.

Now we consider the scalar field equation. Substituting the fields 
\eqref{fs:fields:Eq}, and the metric \eqref{fs:metric:Eq} 
into the equation of motion for the scalar field \eqref{fs:scalar:Eq}, 
we obtain
\Eqr{
2(p-1)\alpha^{-1}\e^{-2A}\left[A''+(D-2)\left(A'\right)^2
-\frac{\alpha^2}{p-1}\left(-\frac{\Lambda}{p-1}
+\frac{f^2}{4}\right)\right]=0\,.
   \label{fs:scalar3:Eq}
}
Combining Eqs. (\ref{fs:Einstein eq:Eq}) and (\ref{fs:scalar3:Eq}), 
the Einstein equations are conveniently equivalent to 
\Eqrsubl{fs:Einstein2 eq:Eq}{
&&\hspace{-1cm}R_{\mu\nu}(\Xsp)-\beta\left(-\frac{\Lambda}{p-1}
+\frac{f^2}{4}\right) q_{\mu\nu}(\Xsp)=0\,,
   \label{fs:Einstein-mn2:Eq}\\
&&\hspace{-1cm}(D-1)A''+\frac{2(p-1)^2}{(D-2)\alpha^2}
\left[(D-2)\left(A'\right)^2-\frac{\alpha^2}{(p-1)}
\left(-\frac{\Lambda}{p-1}
+\frac{f^2}{4}\right)\right]=0\,,
   \label{fs:Einstein-yy2:Eq}\\
&&\hspace{-1cm}R_{ab}(\Zsp)
-\left[\beta\left(-\frac{\Lambda}{p-1}
+\frac{f^2}{4}\right)+\frac{f^2}{2}\right]\gamma_{ab}(\Zsp)=0\,,
   \label{fs:Einstein-ab2:Eq}
}
where the constant $\beta$ is defined by
\Eq{
\beta=\frac{\alpha^2}{p-1}-\frac{2(p-1)}{D-2}\,.
   \label{fs:beta:Eq}
}
From Eqs.(\ref{fs:scalar3:Eq}) and (\ref{fs:Einstein-yy2:Eq}), 
we get 
\Eq{
A(y)=\ell\left(y-y_0\right)\,,
}
where $y_0$ is constant, and $\ell$ is defined by
\Eq{
\ell=\pm\,\alpha\,\sqrt{\frac{1}{(p-1)(D-2)}\left(-\frac{\Lambda}{p-1}
+\frac{f^2}{4}\right)}\,.
}
If we choose $\alpha$ so that $\beta>0$\,, it follows that
\Eq{
\alpha>\sqrt{\frac{2}{D-2}}\,(p-1)\,,~~~~~~
\alpha<-\sqrt{\frac{2}{D-2}}\,(p-1)\,.
}
Hence, the internal space Z has the positive curvature which 
is clear from an inspection of \eqref{fs:Einstein-ab2:Eq}.
So the field equations lead to the $D$-dimensional metric 
\Eq{
ds^2=\e^{2\ell(y-y_0)}\left[-dt^2+\e^{2Ht}
\delta_{mn}dx^{m}dx^{n}
+dy^2+\gamma_{ab}(\Zsp)dz^adz^b\right]\,,
  \label{fs:D-metric2:Eq}
}
where the Hubble parameter $H$ is given by 
\Eq{
H^2=\frac{\beta}{n-1}\left(-\frac{\Lambda}{p-1}+\frac{f^2}{4}\right)\,.
\label{heq}
}
In the limit of $H\rightarrow 0$, we see that the Ricci tensor 
of the Z space leads to $R_{ab}(\Zsp)\rightarrow
\frac{1}{2}f^2\gamma_{ab}(\Zsp)$\,.
As the internal space is essentially 
supported by the field strength,  we can keep the 
geometrical property of the internal space in the limit 
$H\rightarrow 0$\,.

Before concluding this subsection, we comment about the ansatz for the field 
strength $F$\,. In this paper, we have simply assumed 
that the indices of non-vanishing components of 
the field strength could be along the internal space Z 
to obtain the de Sitter spacetime for warped compactifications. 
If the constant field strength $F$ 
has the components along our $(n+1)$-dimensional spacetime M, the Ricci 
tensor on M is proportional to the $(n+1)$-dimensional metric with negative 
sign that is no longer de Sitter spacetime.

\subsection{Brane world models}
 
Now we 
construct 
cosmological models directly from our solution
\eqref{fs:D-metric2:Eq},
following the brane world approach.
Since the details are discussed in Ref. \cite{Minamitsuji:2011gn},
we explain only the essence of our model.
The $D$-dimensional metric  
\eqref{fs:D-metric2:Eq}
 is given in terms of the warped product of 
the non-compact direction $y$,
the $(D-n-1)$-dimensional compact space $\Zsp$
and the $n$-dimensional de Sitter spacetime $\Xsp$
 In Ref. \cite{Minamitsuji:2011gn}, we fix $n=4$).
To realize the finite bulk volume,
we cut the spacetime along $y=y_0$,
where $y_0$ is a constant. 
Then, the original spacetime is glued to its identical copy
across the codimension-one discontinuity at $y=y_0$,
which is now identified as the $(D-1)$-dimensional brane world
composed of the $\Xsp$ and $\Zsp$ spaces.
Although our construction 
is very similar to the five-dimensional Randall-Sundrum (RS) model \cite{rs},
a manifest difference is that our brane world
still contains the compact internal space $\Zsp$.
Thus after integrating over $\Zsp$ on the brane,
we can identify the $n$-dimensional de Sitter space $\Xsp$ 
as our $n$-dimensional Universe
with a finite effective 4D Planck mass.
It also should be noted that the inclusion of the
gauge field strength acting on the $\Zsp$ space is necessary
to obtain 
a finite size of the compact internal space
in the limit of the vanishing de Sitter expansion rate,
which is the realization of the 
modulus stabilization of the $\Zsp$ space
in the brane world.

We then discuss the spectrum
of the tensor metric perturbations
realized in our four-dimensional de Sitter universe.
In Ref. \cite{Minamitsuji:2011gn},
we showed
the existence of the three kinds of eigen modes realized
in our Universe,
i.e., 
the massless zero mode,
the continuum of Kaluza-Klein (KK) modes
associated with the noncompact direction $y$
and the discrete KK modes
arising due to compactification of $\Zsp$ on the brane.
The zero mode would reproduce the 4D gravity on the brane,
and 
all the KK excitations 
are heavier than the critical mass
in the de Sitter space,
above which the wavefunction of the corresponding mode
exhibits a damped oscillation.
Thus
our four-dimensional de Sitter Universe
is free from any massive excitation \cite{Minamitsuji:2011gn}.
We also showed that
both 
the original model found in \cite{Neupane:2010ya}
and 
our model
suffer an instability 
against the scalar perturbations,
associated with the decompactification in the $y$ direction,
although in our model
the internal space $\Zsp$
can be fixed by the field strength. 
We expect that
Casimir effects of the bulk fields
would stabilize the $y$ space,
and this subject is under active study \cite{mu}.

In the rest of the paper, 
we will present another construction of the 
cosmological models based on the lower-dimensional effective theory,
integrating over the $\Zsp$ space.
In the second approach, 
the noncompact $y$-direction should be regarded
as an external direction in our Universe
in contrast to the brane world approach.

\section{Lower-dimensional effective theory}
  \label{sec:ef}
In this section, 
we will consider another construction of 
our de Sitter Universe 
as well as
fixing the moduli in the lower-dimensional effective theory,
following the ordinary compactification approach.
For simplicity, we consider 
the internal moduli degrees of freedom of the metric of 
internal space Z in the present paper.

  

Now, we assume the $D$-dimensional metric 
\Eqr{
&&ds^2=\e^{2A(v)}\left[w_{PQ}(\Msp)dv^Pdv^Q
+\e^{2\psi(v)}\gamma_{ab}(\Zsp)dz^adz^b\right]\,,
   \label{fe:metric:Eq}
}
where $\psi(v)$ is the moduli degrees of freedom on the Z space, and 
$w_{PQ}(\Msp)$ is the $(n+1)$-dimensional metric, and 
$\gamma_{ab}(\Zsp)$ is the metric of the $(D-n-1)$-dimensional Einstein 
space $\Zsp$. 
 
In contrast to the previous section, $\Msp$ contains the 
direction of the infinite line $y$
in the previous sections,
which is regarded as an external direction
in the effective theory approach.
Furthermore, we assume that the scalar fields are expressed as in 
\eqref{fs:scalar2:Eq}, and 
$F$ is a fixed by \eqref{fs:strength2:Eq}. 
Hence, the moduli $\psi$ and the function $A$  
are the only dynamical variables in the effective theory. 
In the following, we construct the $(n+1)$-dimensional effective 
theory after compactifying the Z space. 
However, 
in contrast to 
the brane world model
constructed from the higher-dimensional solution
\eqref{fs:D-metric2:Eq}
where the $n$-dimensional spacetime X becomes our
external directions,
now the $(n+1)$-dimensional 
spacetime $\Msp$
including the noncompact direction $y$
is assumed to be our external directions. 
The $(n+1)$-dimensional effective action for these variables 
can be obtained by evaluating the $D$-dimensional action 
\eqref{fs:action:Eq}. First, for the metric 
\eqref{fe:metric:Eq}, the $D$-dimensional scalar curvature 
$R$ is expressed as
\Eqr{
R&=&\e^{-2A}\left[R(\Msp)+\e^{-2\psi}R(\Zsp)
-2(n-1)\triangle_{\Msp}A+2p\lap_{\Msp}\psi
    +p(p+1)w^{PQ}\pd_P\psi\pd_Q\psi\right.\nn\\
&&   \left.+\left\{(n-1)(p-1)+p(D-1)\right\}w^{PQ}\pd_PA\pd_QA
   +2p(D-1)w^{PQ}\pd_PA\pd_Q\psi\right]\,,
   \label{fe:Ricci:Eq}
}
where $\triangle_{\Msp}$ and $\triangle_{\Zsp}$ are 
the Laplace operators on the space of 
$\Msp$ and the space $\Zsp$, and 
$R_{PQ}(\Msp)$ and $R_{ab}(\Zsp)$ are the Ricci tensors
of the metrics $w_{PQ}$ and $\gamma_{ab}$, respectively. 
We assume that the internal space $\Zsp$ satisfies the condition 
$R_{ab}(\Zsp)=\lambda \gamma_{ab}(\Zsp)$, 
where $\lambda$ characterizes the internal space curvature.
Here $\beta$ is given by (\ref{fs:beta:Eq})\,. 
Inserting the Eqs.~(\ref{fs:fields:Eq}) and (\ref{fe:Ricci:Eq}) 
 into \eqref{fs:action:Eq}, we get 
\Eqr{
S&=&\frac{1}{2\tilde{\kappa}^2} \int_{\Msp} \e^{(D-2)A+p\psi} 
\left[\left\{R(\Msp)+p\lambda \e^{-2\psi}-2\Lambda
-\frac{f^2}{2}\right\}\ast_{\Msp}{\bf 1}_{\Msp}
-2(n-1)d\ast_{\Msp}dA\right.\nn\\
&& \hspace{-0.5cm}-2pd\ast_{\Msp}d\psi  
-\left\{(n-1)(p-1)+p(D-1)\right\}dA\wedge \ast_{\Msp} dA
   -2p(D-1)dA\wedge \ast_{\Msp}d\psi\nn\\
&&   \left.   \hspace{-0.5cm}-p(p+1)d\psi\wedge \ast_{\Msp}d\psi
\right]\,,
   \label{fe:d-action:Eq}
}
where $\ast_{\Msp}$ is the Hodge operator on the M space 
and $\tilde{\kappa}$ is given by 
$\tilde{\kappa}=V^{-1/2}\kappa$ with the volume of the internal space Z 
\Eq{
V\equiv\int_{\Zsp}\ast_{\Zsp}{\bf 1}_{\Zsp}\,.
  \label{fe:volume:Eq}
} 
Here $\ast_{\Zsp}$ is the Hodge operator on the Z space. 

Using the conformal transformation 
$w_{PQ}(\Msp)=\e^{-2[(D-2)A+p\psi]/(n-1)}w_{PQ}(\bMsp)$, 
the $(n+1)$-dimensional action 
\eqref{fe:d-action:Eq} is expressed in terms of the 
variables in the Einstein frame as 
\Eqr{
S&=&\frac{1}{2\tilde{\kappa}^2} \int_{\bMsp}
 \left[\left\{R(\bMsp)-V\left(\bar{A},~\bar{\psi}\right)\right\}
 \ast_{\bMsp}{\bf 1}_{\bMsp}
 -\frac{1}{2}d\bar{A}\wedge\ast_{\bMsp}d\bar{A}\right.\nn\\
 &&\left.
 -\frac{1}{2}\frac{c_2}{\sqrt{c_1c_3}}d\bar{A}\wedge\ast_{\bMsp}d\bar{\psi}
 -\frac{1}{2}d\bar{\psi}\wedge\ast_{\bMsp}d\bar{\psi}\right],
   \label{fe:Ed-action:Eq}
}
where 
$R(\bMsp)$ is the Ricci scalar with respect to the metric 
$w_{PQ}(\bMsp)$, and we have dropped the surface terms coming from 
$\lap_{\bMsp}A$, $\lap_{\bMsp}\psi$, and the 
potential $V\left(\bar{A},~\bar{\psi}\right)$, fields $\bar{A}$, 
$\bar{\psi}$, the constants $c_i~(i=1, 2, 3)$ are defined by 
\Eqrsubl{fe:c:Eq}{
V\left(\bar{A}, \bar{\psi}\right)&=&\exp\left[-\frac{2(D-2)\bar{A}}
{(n-1)\sqrt{c_1}}\right]\left[
2\Lambda\exp\left\{-\frac{2p\bar{\psi}}{(n-1)\sqrt{c_3}}\right\}\right.\nn\\
&&\left.+\frac{f^2}{2}
\exp\left\{-\frac{2np\bar{\psi}}{(n-1)\sqrt{c_3}}\right\}
-p\lambda\exp\left\{-\frac{2(D-2)\bar{\psi}}{(n-1)\sqrt{c_3}}\right\}
\right]\,,
     \label{fe:potential:Eq}\\
\bar{A}&=&\sqrt{c_1}A\,,~~~~~~ 
\bar{\psi}=\sqrt{c_3}\psi\,,\\
c_1&=&2\left[\frac{n}{n-1}(D-2)-2(D-1)\right](D-2)\nn\\
&&+2\left[n-1+\frac{2}{\alpha^2}(p-1)\right](p-1)+2p(D-1)\,,\\
c_2&=&\frac{4(D-2)p}{n-1}\,,\\
c_3&=&2p\left(\frac{n-1}{p}+1\right)\,.
}
Here $\lap_{\bMsp}$ is the Laplace operator 
constructed from the metric $w_{PQ}(\bMsp)$. 
The form of the potential \eqref{fe:potential:Eq} 
implies that the warp factor $\bar{A}$ cannot be fixed by the background 
fields. This is consistent with the fact that the analysis of the 
the scalar perturbations for the solutions which are discussed in this 
paper is unstable. We have explained these in 
\cite{Minamitsuji:2011gn}. 

\begin{figure}[h]
 \begin{center}
  \includegraphics[keepaspectratio,scale=1.0]{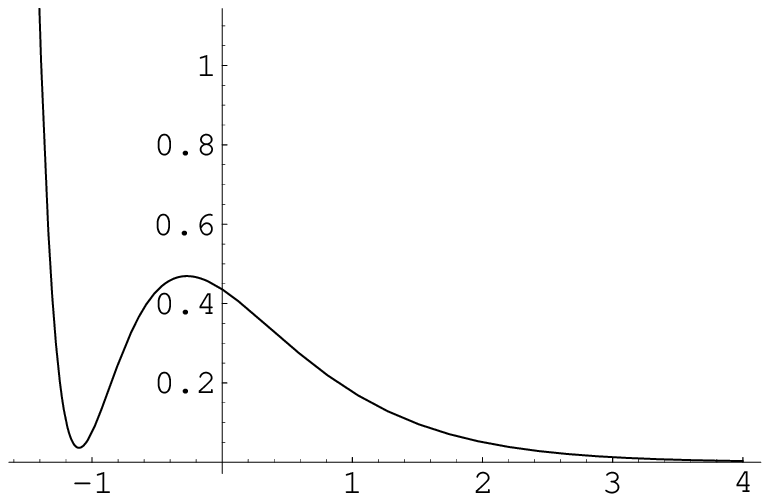}
\put(-260,160){$V(1,~\bar{\psi})/\Lambda$}
\put(10,10){$\bar{\psi}$}
  \caption{The moduli potential given in 
  (4.6a) is depicted. 
  We set $n=3$, $p=6$, $\bar{A}=1$, 
  $f=0.17$, $\lambda=0.229$ and $\alpha=0.5$ in the unit of $\Lambda=1$.
  The minimum of the potential is 
  located at $\bar{\psi}_0=-1.1000$ and its value is 
  $\Lambda^{-1}V\left(1, \bar{\psi}_0\right)=0.0365$\,. 
  If we fix the value of the function $\bar{A}$,  
  we get the small energy 
  value of the moduli potential at the local minimum.}
  \label{fe:moduli:fig}
 \end{center}
\end{figure}

\begin{figure}[h]
 \begin{center}
  \includegraphics[keepaspectratio, scale=1.0]{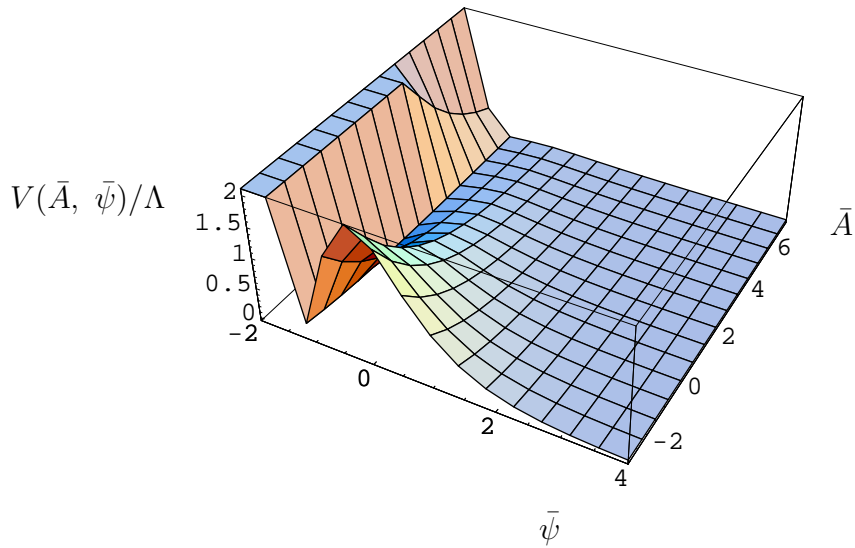}
\put(-300,110){$V(\bar{A},~\bar{\psi})/\Lambda$}
\put(-100,-15){$\bar{\psi}$}
\put(10,100){$\bar{A}$}
\caption{The moduli potential given in 
(4.6a) is depicted. We set $n=3$, $p=6$, $f=0.17$, $\lambda=0.229$ and $\alpha=0.5$ in the unit of $\Lambda=1$. We can find the minimum of the potential with respect to the field $\bar{\psi}$ while there is not local minimum of the potential for the direction of $\bar{A}$.}
  \label{fe:3d:fig}
 \end{center}
\end{figure}

Then, in the following, we fix the value of $\bar{A}$ by
assuming some additional stabilization mechanism which does not affect
the dynamics of the other moduli $\bar{\psi}$,
and consider the stabilization of 
$\bar{\psi}$. 
In order to fix the moduli degrees of freedom $\bar{\psi}$, 
the moduli potential has to have a minimum or at least a local minimum. 
The moduli potential energy at the minimum is given by 
\Eqr{
V\left(\bar{A}, \bar{\psi}_0\right)&=&
\exp\left[-\frac{2(D-2)\bar{A}}{(n-1)\sqrt{c_1}}\right]
\left[-\frac{1}{2}(n-1)f^2
\exp\left\{-\frac{2pn\bar{\psi}_0}{(n-1)\sqrt{c_3}}\right\}
\right.\nn\\
&&\left.\hspace{-1cm}
-
\lambda(D-p-2)
\exp\left\{-\frac{2(D-2)\bar{\psi}_0}{(n-1)\sqrt{c_3}}\right\}
\right]\,.
    \label{fe:minimum:Eq}
}
In the present model, setting $n=3$, $p=6$, $\bar{A}=1$, 
$f=0.17$, $\lambda=0.229$ and $\alpha=0.5$ in the unit of $\Lambda=1$, 
we find the stable minimum point $\bar{\psi}_0=-1.1000$,  where 
$\Lambda^{-1}
V\left(1, \bar{\psi}_0\right)=0.0365$\,. Since the potential 
energy is proportional to the warp factor $\e^{-\bar{A}}$, 
the $(n+1)$-dimensional 
effective cosmological constant can be dropped exponentially as 
$\bar{A}$ increases. 
Hence, if we choose the value of the function $\bar{A}$ appropriately, 
we get the small energy value of the potential 
at $\bar{\psi}=\bar{\psi}_0$. 
We illustrate the moduli potential in Fig. \ref{fe:moduli:fig} and 
\ref{fe:3d:fig}.

In order to give the positive value of the potential energy, 
the constant $\bar{\psi}_0$ should 
satisfy the condition 
\Eq{
\exp\left[-\frac{pn\bar{\psi}_0}{(n-1)(D-2)\sqrt{c_3}}
\right]<
\frac{2}{(n-1)f^2}
\lambda
(D-p-2)\,.
}

In the gravity system with a positive cosmological constant and finite 
volume of the internal space, upon setting $D>p+2$, 
a positive curvature term of the internal space gives
a dominant contribution to the moduli potential at small 
$\bar{\psi}$, while a positive cosmological constant term becomes dominant
for large $\bar{\psi}$. Hence the moduli potential is unbounded
from below as $\bar{\psi}\rightarrow -\infty$ and 
drops exponentially as $\bar{\psi}\rightarrow\infty$.
Then the internal space Z either shrinks to zero volume
or is decompactified. We also obtain the run away type moduli potential 
even if $\Lambda<0$\,. 
However, if we include the field strength with a cosmological constant, 
we can find a stable minimum for the moduli potential. 
The cosmological constant and the field strength force to expand the 
extra dimension while the curvature of the internal spacetime 
forces to contract it. These combinations produce a local minimum of the
effective potential. Hence the role of the field strength is 
important to find a minimum of the potential \cite{Carroll:2001ih}.  
The potential energy at the minimum is equivalent to 
the $(n+1)$-dimensional cosmological constant. Then we can obtain 
the stable in the dS${}_{n+1}\times$Z background. 
If the universe may be created one the top of the potential
hill, the universe rolls down to the potential minimum 
whose value is positive. For other choice of the parameters, 
we can also have a negative potential minimum. Then the universe
evolves into the static AdS${}_{n+1}\times$Z, which is a 
stable spacetime. 
As the moduli potential without the cosmological constant cannot provide 
the positive value at the local minimum even if we have a field 
strength, the contribution of the cosmological constant is essential 
to derive de Sitter compactifications. 

\section{Discussions}
  \label{sec:Discussions}
In the first part of this paper, 
we have discussed the warped de Sitter compactifications
 in higher-dimensional theory. 
We have obtained 
warped de Sitter spacetimes
with a positively or negatively curved internal space
to investigate the possible generalizations
of the solutions found in \cite{Neupane:2010ya}.
We also 
presented the $n$-dimensional warped de Sitter compactification  
due to the contribution of the matter fields.  
These solutions give the accelerating expansion of 
the $n$-dimensional universe.
The construction of the cosmological model
is directly obtained from our solutions,
following the brane world approach
(see \cite{Minamitsuji:2011gn} for details).
The solutions without field strength were obtained by the product 
type form of the warp factor. 
This structure requires that warp factor of the metric is expressed as 
the form $A_0(x)A_1(y)$, where $A_0(x)$ is a function of the 
four-dimensional coordinates $x^{\mu}$, and $A_1(y)$ is
a function on the internal space. In the pure gravity system, as 
the function $A_0(x)$ in the metric describes the conformal rescaling of 
the $n$-dimensional metric, we could set $A_0=0$ without loss of generality. 
Supposing that our universe stays at a constant position 
in the internal space Y, we have shown that the evolution of  
$n$-dimensional universe gives an $n$-dimensional de Sitter space or 
FRW universe. The solutions in the $D$-dimensional theory give 
the $n$-dimensional de Sitter spacetime for the case of 
not only the pure gravity system field but also the gravity with matter fields 
which depend only on the coordinates of $y$. 
However, if the scalar field 
depends only on the coordinate of the $n$-dimensional spacetime, 
the solution of field equations denotes that the power of the scale factor 
of $n$-dimensional spacetime is too small 
to give a realistic expansion law for $n$-dimensional FRW universe. 
For the compactification with field strength, we have simply required 
that non-vanishing components of the field strength should 
be along the internal space $\Zsp$. 
An important consequence of this is 
obtaining the de Sitter spacetime  
from warped compactifications. If the constant field strength $F$ 
goes through our $n$-dimensional spacetime $\Xsp$, the Ricci 
tensor on $\Xsp$ is proportional to the $n$-dimensional metric with negative 
sign that is no longer de Sitter spacetime. 
Hence we were only concerned with the field strength along Z.

In the second part of this paper, 
we have presented another 
construction of a de Sitter spacetime 
in terms of the $(n+1)$-dimensional effective 
theory after integrating over the $\Zsp$ space.
The $(n+1)$-dimensional spacetime $\bMsp$
was regarded as 
our Universe. 
We have calculated the moduli potential 
and discussed its stability using the 
moduli potential. 
Assuming some stabilization mechanism of the volume moduli
which does not affect the dynamics of the internal space moduli
(for example, via Casimir effects of the bulk fields
 \cite{mu}),
the
cosmological constant and field strengths force the internal space
to expand,
while the curvature of the internal spacetime 
forces it to contract.
These combination produces a local minimum of the
moduli potential. 
In the $(n+1)$-dimensional effective theory, 
the scale of the compact internal space $\Zsp$
is stabilized by 
balancing the gauge field strength wrapped
around the internal space
and the curvature term of the internal space 
with the cosmological constant. 
For some choices of the parameters,
these contributions could produce a local minimum 
of the effective potential with a positive value 
which corresponds to the dS${}_{n+1}\times\Zsp$ 
background in the effective theory, while the solution in the 
original higher-dimensional gravitational theory gave the geometry of 
dS${}_n\times\mathbb{R}\times\Zsp$.
If the $(n+1)$-dimensional universe is created near the top of 
the potential hill, the moduli rolls down the potential hill 
and finds a stable minimum point. Since the moduli potential eventually 
turns out to be positive or negative, the $(n+1)$-dimensional 
background geometry becomes dS${}_{n+1}$ or AdS${}_{n+1}$ spacetime. 
As the moduli potential without the cosmological constant leads to  
the AdS${}_{n+1}$ at the local minimum even if we have a field 
strength, we need the contribution of the cosmological constant 
to obtain de Sitter compactifications. 

The next step may be to develop a framework to understand 
generalizations of warped compactifications in which 
one varies the matter fields or the boundary conditions or other
details though this program 
has not been actually achieved in the present paper. 
Along this way, we will clarify some properties of cosmic acceleration 
for the warped compactification elsewhere.

\section*{Acknowledgments}
K.U. is supported by Grant-in-Aid for 
Young Scientists (B) of JSPS Research, under Contract No. 20740147.



\end{document}